\documentstyle{elsart}

\input epsf

\def\Journal#1#2#3#4{{#1} {\bf #2}, #3 (#4)}

\def\NIM{\em Nucl. Instrum. Methods}

\def\PLB{{\em Phys. Lett.}  B}

\def\PRD{{\em Phys. Rev.} D}

\def\PR{\em Phys. Rep.}

\begin{document}
\begin{frontmatter}
\title{Test of CsI(T$\ell$) crystals for the Dark Matter Search }

\author{H.J.Kim\thanksref{address}\thanksref{LSU},}
\author{H.J.Ahn, S.K.Kim, E.Won\thanksref{KEK},} 
\author{T.Y.Kim}
\address{Department of Physics, Seoul National University, Seoul 151-742, Korea} 

\author{Y.D.Kim}
\address{Department of Physics, Sejong University, Seoul 143-747, Korea}

\author{M.H.Lee}
\address{KEK, Tsukuba, Ibaraki 305-0801, Japan}

\author{J.S.Chai, J.H.Ha\thanksref{KAERI}} 
\address{Korea Cancer Center Hospital, Seoul, Korea}

\thanks[address] {Corresponding author; E-mail: hjkim@hep1.snu.ac.kr; Tel: +82 2 876 2801; FAX: +82 2 875 4719}
\thanks[LSU] {Also affiliated with Department of Physics and Astronomy, Louisiana State University, Baton Rouge, LA 70803, USA}
\thanks[KEK] {Also affiliated with KEK, Tsukuba, Ibaraki 305-0801, Japan}
\thanks[KAERI] {Present address: Korea Atomic Energy Research Institute, Taejon, 305-600, Korea}

\begin{abstract}   
Searches for weakly interacting massive particles(WIMP) can be based on the detection
of nuclear recoil energy in CsI(T$\ell$) crystals.
We demonstrate that low energy gamma rays down to few keV is detected with
CsI(T$\ell$) crystal detector. A clear peak at 6 keV is observed using X-ray
source.  Good energy resolution and linearity have been achieved down to X-ray
region.
In addition, we also show that alpha particles and gamma rays can be
clearly separated using the different time characteristics of the crystal.
\end{abstract}


\end{frontmatter}

\section{Introduction}

Several evidences from a variety of sources indicate that the universe contains a 
large amount of dark matter~\cite{dm0}. The most strong evidence for the existence of 
dark matter comes from the galactic dynamics. There is simply not enough luminous
matter observed in spiral galaxies to account for the observed rotational 
curves~\cite{dm1}.
Among several dark matter candidates, one of the most prominent candidate is 
the weakly-interacting massive particles(WIMP).
The leading WIMP candidate is perhaps the neutralino, 
the lightest super-symmetric particles such as photinos, 
Higgsinos and Z-inos~\cite{susy}.
These particles typically have masses between 10 GeV and a few TeV 
and couple to ordinary matter only with weak interactions.  
The elastic scattering of WIMP with target nuclei 
could be detected by measuring the recoil energy of the nucleus, 
which is up to several tens of keV~\cite{witten}. 
Recently, a great deal of attention has been
drawn to crystal detectors since the detection technique is already 
developed and radioactive background from the crystal is under control.
Especially, the most stringent limit
for the direct detection of WIMP has been
established using the NaI(Tl) crystal detector~\cite{NaI}. 
They achieved as low threshold as 6 keV and
relatively good separation between the recoiling events and the ionizing
events by background $\gamma$'s using the difference of the 
scintillation decay time. 

Recently, positive signal of annual modulation has been reported by DAMA
group~\cite{signal}.  Looking at the similar sensitivity region with 
other experiments which involves different systematics is absolutely necessary 
to confirm their results.
It has been noted by several authors that CsI(T$\ell$) crystal may give better
performance for the separation between recoiling events and the ionizing
events by background $\gamma$ ~\cite{sep}.
Although the light yield of CsI(T$\ell$) crystal
is slightly lower than NaI(Tl) crystal, better particle separation can be more
advantageous for WIMP search.   Also CsI(T$\ell$) has much less hygroscopicity than NaI(Tl), and has higher density (see Table I).
The spin-independent cross section of WIMP is larger for 
CsI(T$\ell$) than NaI(Tl) because CsI(T$\ell$) has a compound with two similar heavy mass nuclei while spin-dependent cross section will be comparable.  
Moreover hundreds of tons of CsI(T$\ell$) 
crystals are  already being used for several detectors in high energy 
experiment~\cite{high}. 
Thus fabricating large amount of crystals is quite feasible.
In this report, we have studied the characteristics of CsI(T$\ell$) 
crystal for the possibility of dark matter search experiment~\cite{we}. 

\section{Experimental Setup}

\begin{figure}
  \begin{center}
     \epsfxsize=5.0in 
     \epsfbox{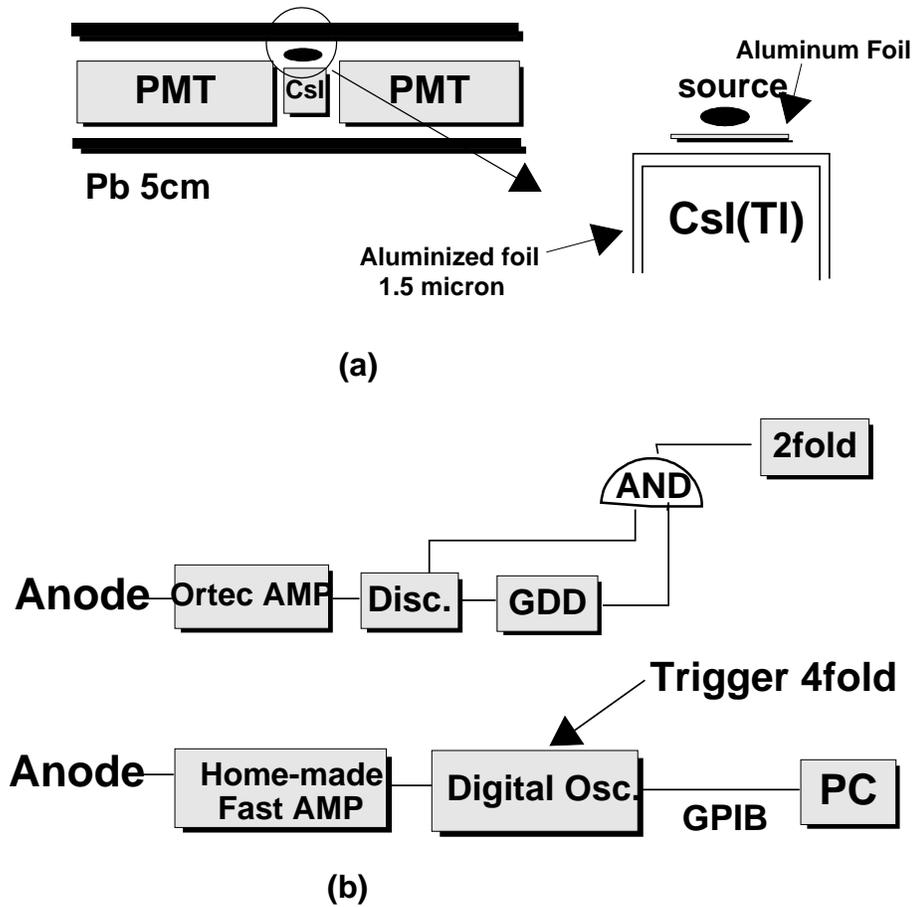}
     \caption{(a) Schematic drawing of the experimental setup 
      and (b) the trigger logic.} 
     \protect\label{setup}
  \end{center}
\end{figure}

We prepared a 3cm$\times$3cm$\times$3cm CsI(T$\ell$) crystal with all surfaces 
polished. 
Photo-multiplier tubes of 2 inch diameter(Hamamtsu H1161) are directly 
attached on two  opposite end 
surfaces. The cathode planes of PMT cover all the area of the crystal
surfaces attached. The other sides are wrapped with 1.5 $\mu$m thick 
aluminized foil window or Teflon tapes followed by black tapes. 
It is necessary
to use only very thin foil for the side where X-ray sources are attached  
that low energy X-rays are not blocked.  For the alpha source, 
additional aluminum foil is located between the aluminized foil and
the source to reduce the $\alpha$ energy. 
 Signals from both PMTs are then amplified using a home-made AMP($\times$8)
with low noise and high slew rate.   Another signals are amplified 
with ORTEC AMP($\times$200) to make the trigger logic.
 Discriminator thresholds are set at the level of single 
photoelectron signal.   By using LED, we confirmed 
that the single photoelectron signal is well above the electronic noise.
In order to suppress the accidental triggers from dark currents,
we delay the signal by 100 ns and then formed a self coincidence for each PMT signal,
which require that at least two photoelectrons occur within 200 ns. 
Then coincidence of both PMT signals are made for the final trigger decision.
In this way the trigger caused by the accidental noises are suppressed
by a great amount. With this condition the effective threshold is four
photoelectrons, which roughly corresponds to 40 photons produced. 
Using the widely accepted light yield of CsI(T$\ell$), $\sim$50,000 photons/MeV, 
our threshold can  be interpreted as 2 keV.
The crystal and PMTs are located inside the of 5 cm
thick lead blocks in order to stop the environmental background. A digital 
oscilloscope is used for the data taking with GPIB interface to a PC with
LINUX system. We developed DAQ system with GPIB and CAMAC interface 
based on the ROOT package~\cite{root} and entire analysis
was performed with the ROOT package too. 
The schematics of the experimental setup and the trigger elements are shown in Figure~\ref{setup} a) and b). The digital oscilloscope we
used for our experiment samples the signal at 1 Gs/sec with 8 bit pulse 
height information and two channels are read out simultaneously. 
Full pulse shape informations are saved for the further analysis.

\section{Calibration}

\begin{figure}
  \begin{center}
     \epsfxsize=5.0in 
     \epsfbox{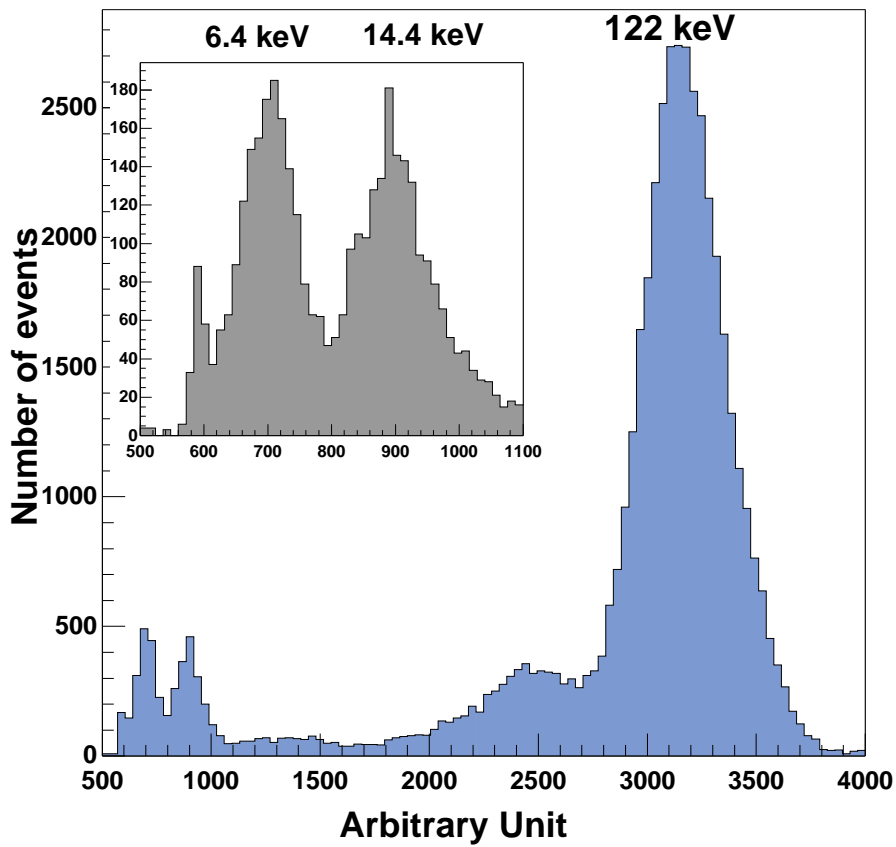}
      \caption{Pulse height spectrum of CsI(T$\ell$) for $^{57}$Co source. 
       The left top plot is zoomed pulse height spectrum of 
       the low energy X-ray.}
      \protect\label{co57}
  \end{center}
\end{figure}

\begin{figure}
  \begin{center}
     \epsfxsize=5.0in 
     \epsfbox{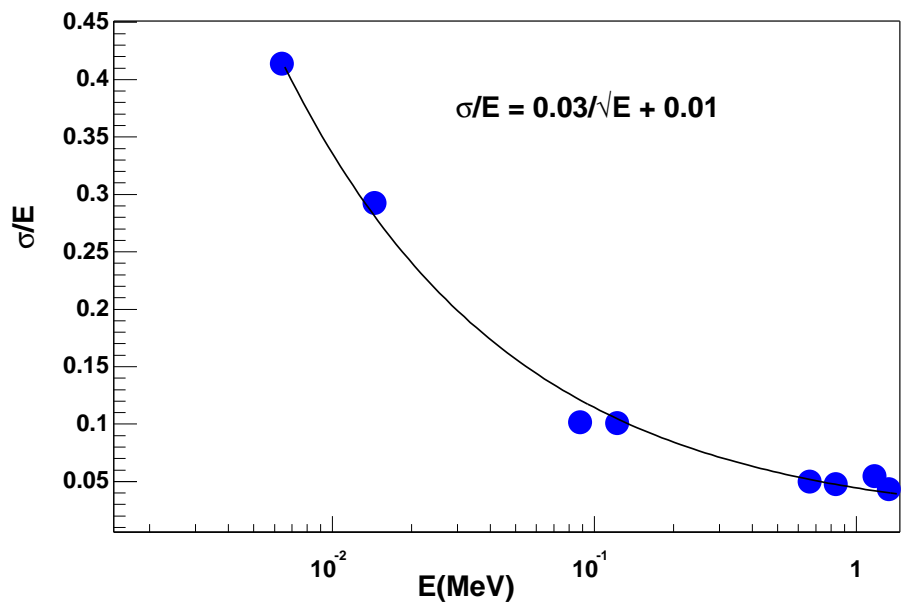}
       \caption{The energy resolution of CsI(T$\ell$) with one-side PMT. 
        The solid curve shows the best fit to data.}
      \protect\label{resol}
  \end{center}
\end{figure}

We have performed measurements of X-rays, $\gamma$-rays, and alpha particles
using various radioactive sources with the setup described in the previous
section. 
The energy spectra of  X-rays and $\gamma$ rays from the $^{57}$Co source 
is given in Fig.~\ref{co57}. The highest peak is from the gamma
ray of 122 keV. Shown in left side of broad distribution of pulses are the Compton
edge.  The energy resolution at 122 keV is about 7\%. Also, the X-ray peak
at 6.4 and 14.4 keV  are clearly seen with energy resolution of 30 and 20\%, respectively.  
This resolution is not much worse than that of NaI(Tl) crystal~\cite{NaI}.  
Many calibration sources such as $^{57}$Co, $^{109}$Cd, $^{137}$Cs, $^{54}$Mn and $^{60}$Co are 
used for the determination of linearity and resolution. 
Fig.~\ref{resol} shows the energy resolution of CsI(T$\ell$) crystal with PMT on 
each side.
The best fit of the resolution with following the parameterization is 
\begin{equation}
\frac{\sigma}{\rm{E(MeV)}} = \frac{0.03}{\sqrt{\rm{E(MeV)}}} \oplus 0.01, 
\end{equation}
and it becomes
\begin{equation}
\frac{\sigma}{\rm{E(MeV)}} = \frac{0.02}{\sqrt{\rm{E(MeV)}}} \oplus 0.01 
\end{equation}
, when we add PMT signals from both sides. 

\begin{figure}
  \begin{center}
     \epsfxsize=5.0in 
     \epsfbox{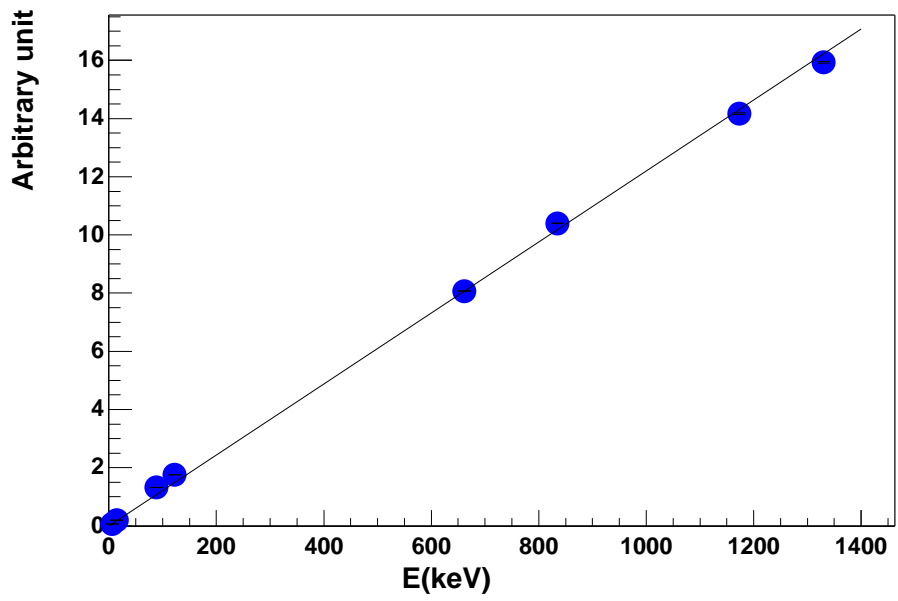}
      \caption{ Linearity distribution of CsI(T$\ell$) crystal 
      with several different photon sources. 
      The solid line shows the linear fit to data.}
      \protect\label{linear}
  \end{center}
\end{figure}

\begin{figure}
  \begin{center}
     \epsfxsize=5.0in 
     \epsfbox{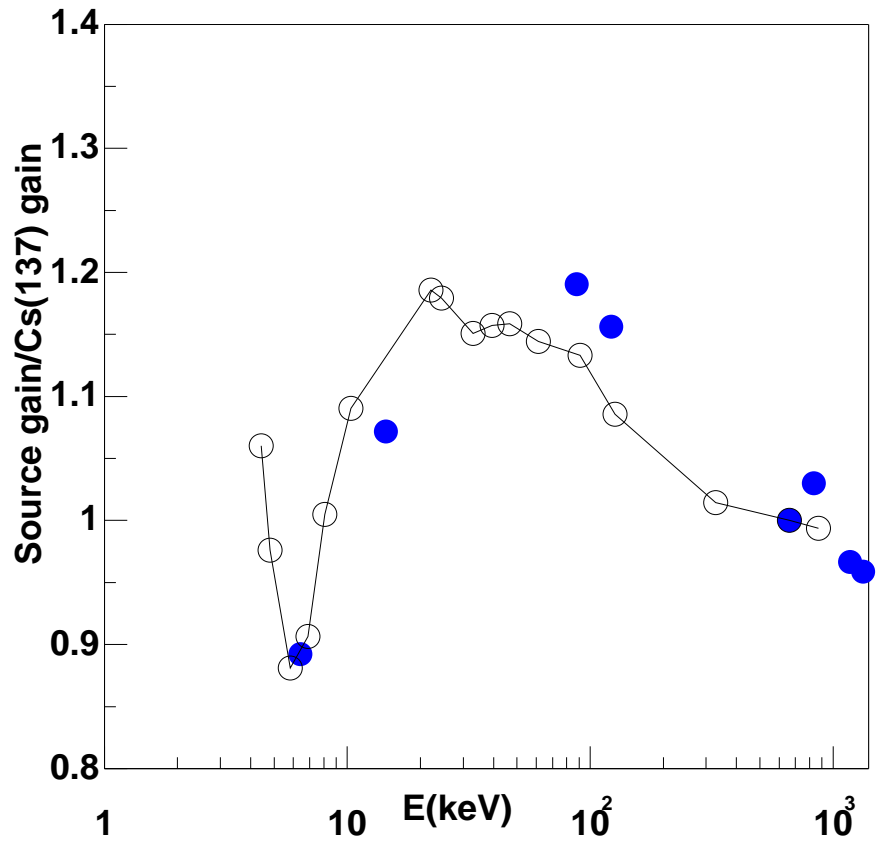}
     \caption{ Response of CsI(T$\ell$) crystal relative to 
     the pulse height of 662 keV gamma ray line from $^{137}$Cs.   
     The filled circles are our data and the open circle with solid lines 
     are the scanned data of 1/8 inch crystal taken from 
     Ref.~\cite{nonlinear}. }  
     \protect\label{nonlinear}
  \end{center}
\end{figure}

The pulse shape is quite linear at high energy as shown in Fig.~\ref{linear} 
but there is  some deviation at low energy as shown in Fig.~\ref{nonlinear}.
The pulse height of the 662 keV $\gamma$-ray line from $^{137}$Cs is defined as unity for the linearity plot.
It turns out that the variation in the response function near the L-, 
K-shell of Cs and I causes nonlinearity at X-ray 
region within 30\%~\cite{nonlinear}.  This is because photoelectrons ejected 
by incident gamma rays just above the K-shell energy have very little kinetic energy 
so that the response drops. Just below this energy, however, K-shell 
ionization is not possible and L-shell ionization takes place. 
Since the binding energy is 
lower, the photoelectrons ejected at this point are more energetic which causes
a rise in the response. 
The pulse shape is linear within 10 \% up to low energy X-ray region if these 
effects are corrected.

\section{Pulse Shape Analysis}

\begin{figure}
  \begin{center}
     \epsfxsize=5.0in 
     \epsfbox{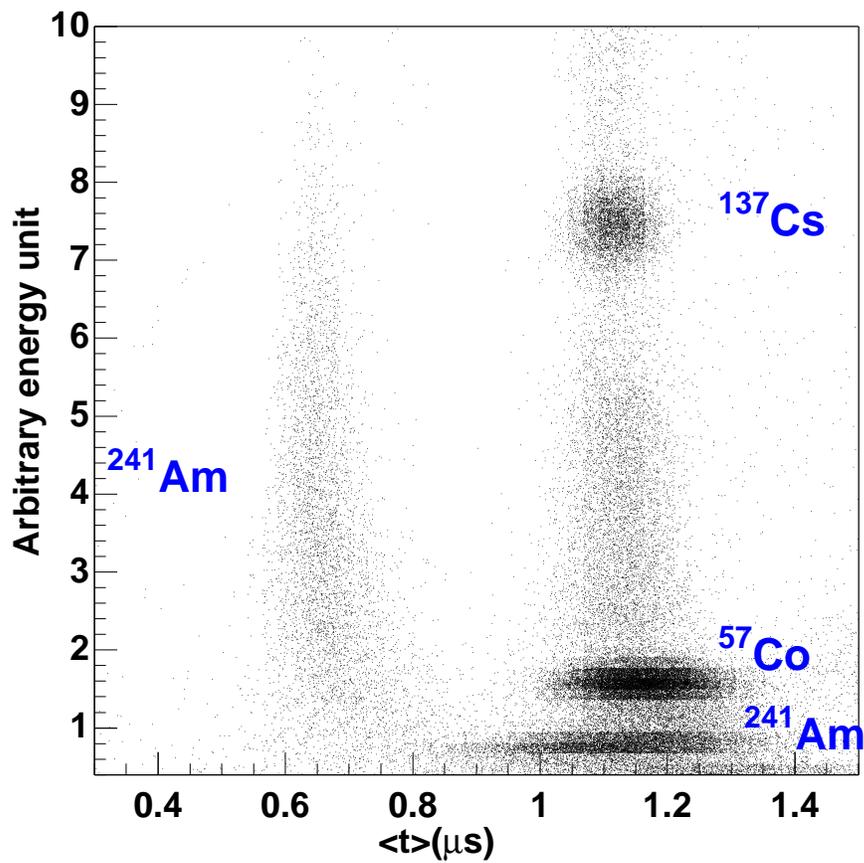}
      \caption{Energy vs. mean time distribution of CsI(T$\ell$) c
       rystal with $^{241}$Am and $\gamma$ sources.}  
      \protect\label{am}
  \end{center}
\end{figure}

\begin{figure}
  \begin{center}
     \epsfxsize=5.0in 
     \epsfbox{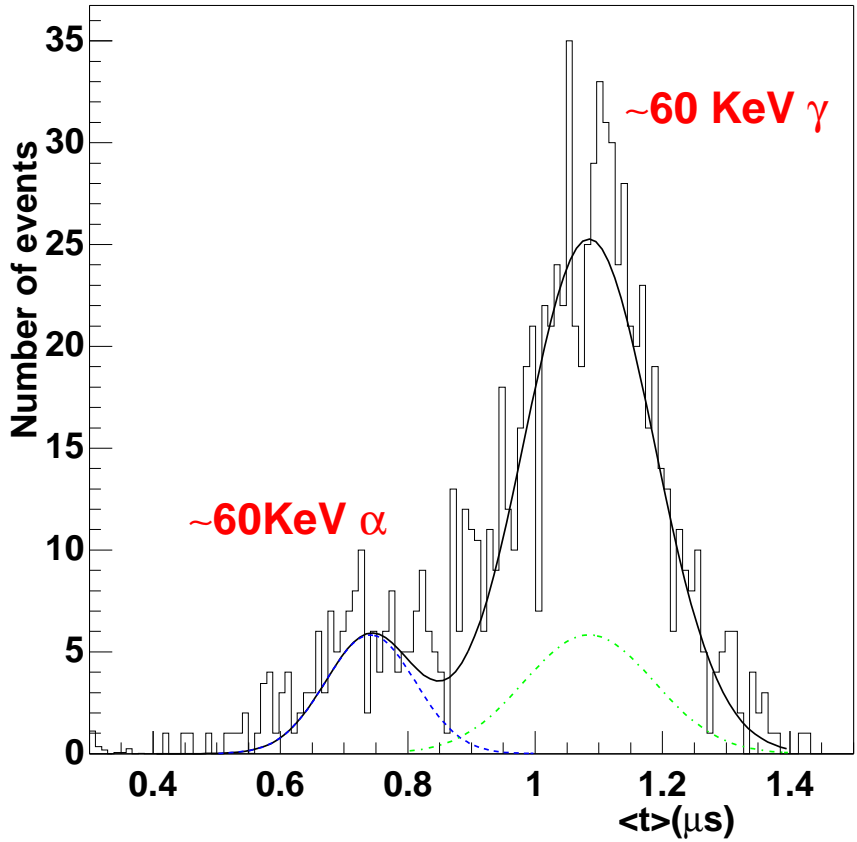}
      \caption{Distribution of the decay time, $<$t$>$, of the 
      CsI(T$\ell$) crystal with $^{241}$Am sources when signals near 
      60 keV are projected. The solid curve shows double Gaussian fit. 
      The dashed Gaussian curve is the decay time of 
      the alpha particle and dotted-dash curve is the decay time of the 
      gamma with the sample pulse height normalized as the alpha's.}
      \protect\label{sep}
  \end{center}
\end{figure}

In many scintillating crystals, electrons and holes produced by ionization are 
captured to form certain meta-stable states and produce slow timing component. 
On the other hand, a larger stopping power from recoiling nucleus produces a higher density of free electrons and holes which favors their recombination into loosely bound systems and results in fast timing component. 
By using this characteristic, we 
may be able to separate X-ray backgrounds from the high ionization loss 
produced by WIMP.
To demonstrate this difference, we measured signals produced by alpha particles 
using $^{241}$Am source. 
Kinetic energy of the alpha particle is 5.5 MeV and the incident energy
was  controlled by the thickness of thin
aluminum foil in front of the crystal. 
Although alpha particle at this energy stops in the crystal, the visible
energy seen by the PMT is about 75\% of the energy. This is due to the
quenching factor for alpha particles and agrees with what were observed
by the other experiments~\cite{alpha}. 
We show the two dimensional histogram
of mean time vs. integrated charge in Fig.~\ref{am}. 
The mean time is 
the pulse height weighted time average, defined as
\begin{equation}
<t> = \frac{\sum t_i \times q_i}{\sum q_i},
\end{equation} 
where $q_i$ is the amplitude of the pulse at the channel time $t_i$ up to 
4 $\mu$s.  It is practically the same as the decay time of the crystal.
Two clear bands in the Fig.~\ref{am} indicate that we can make good separation between the alpha particle and X-ray.
The low energy of X-ray from the $^{241}$Am source is 60 keV.  
In Fig.~\ref{sep}, we projected signals near 60 keV region to the mean time
axis and it shows that the  
decay time for alpha particles is $\sim$700 ns while for X-rays $\sim$1100 ns.
Two peaks are well separated by more than  3 sigma in this energy region.

\section{Conclusion}

We demonstrated that CsI(T$\ell$) crystal can be used to measure low energy 
gamma rays down to few keV. Linearity within 10\% and
good energy resolution have been obtained down to 6 keV X-ray region.
In addition, a good separation of alpha particles
from gamma rays has been achieved by using mean time difference. 
If recoiled ions in the crystal 
behave similar to alpha particles, the mean time difference  would be very 
useful to differentiate WIMP signals  from backgrounds. 
The background study and neutron response on CsI(T$\ell$) study are underway. If this study is successful, a pilot experiment 
with a large amount crystals will be launched in near future.

\begin{ack}
This work is supported by Korean Ministry of Education under 
the grant number  BSRI 1998-015-D00078.
Y.D. Kim wishes to acknowledge the financial support of the Korean
Research Foundation made in the program year of 1998.
\end{ack}

\clearpage

\pagebreak


\begin{table}
\begin{center}
\caption{Comparison of CsI(T$\ell$) and NaI(Tl) characteristics.}\label{tab1}
\vspace{0.5cm}
\begin{tabular}{|l|c|c|} 
\hline 
Property & CsI(T$\ell$) & NaI(Tl) \\
\hline
Density(g/cm$^3$) & 4.53 & 3.67 \\
\hline
Decay constant(ns) & $\sim$1000 & $\sim$250 \\
\hline
Peak emission(nm)  & 550   & 415 \\
\hline
Light yield(relative) & 85 & 100 \\
\hline
Hygroscopicity       & slight & strong \\
\hline
\end{tabular}
\end{center}
\end{table}
\clearpage


\clearpage

\clearpage

\clearpage

\clearpage

\clearpage

\clearpage

\end{document}